# Visible and near infra-red up-conversion in $Tm^{3+}/Yb^{3+}$ co-doped silica fibers under 980 nm excitation

D. A. Simpson[1,†], W. E. K Gibbs[2], S. F. Collins[1], W. Blanc[3], B. Dussardier[3], G. Monnom[3], P. Peterka[4] and G. W. Baxter[*,1]

[1] *Centre for Telecommunications and Micro-Electronics, Optical Technology Research Laboratory, Victoria University, Victoria, 8001, Australia*
[2] *Centre for Atom Optics and Ultrafast Spectroscopy, Swinburne University of Technology, Victoria, 3122, Australia*
[3] *Laboratoire de Physique de la Matière Condensée UMR6622, Université de Nice - Sophia Antipolis Parc Valrose, 06108 Nice, Cedex 2, France*
[4] *Institute of Photonics and Electronics, Academy of Sciences of the Czech Republic, v.v.i. Chaberska 57, 18251 Praha 8, Czech Republic*
† *Now at Quantum Communications Victoria, School of Physics, University of Melbourne, Victoria, 3010, Australia*
*\*Corresponding author: gregory.baxter@research.vu.edu.au*

**Abstract:** The spectroscopic properties of $Tm^{3+}/Yb^{3+}$ co-doped silica fibers under excitation at 980 nm are reported. Three distinct up-conversion fluorescence bands were observed in the visible to near infra-red regions. The blue and red fluorescence bands at 475 and 650 nm, respectively, were found to originate from the $^1G_4$ level of $Tm^{3+}$. A three step up-conversion process was established as the populating mechanism for these fluorescence bands. The fluorescence band at 800 nm was found to originate from two possible transitions in $Tm^{3+}$; one being the transition from the $^3H_4$ to $^3H_6$ manifold which was found to dominate at low pump powers; the other being the transition from the $^1G_4$ to $^3H_6$ level which dominates at higher pump powers. The fluorescence lifetime of the $^3H_4$ and $^3F_4$ levels of $Tm^{3+}$ and $^2F_{5/2}$ level of $Yb^{3+}$ were studied as a function of $Yb^{3+}$ concentration, with no significant energy back transfer from $Tm^{3+}$ to $Yb^{3+}$ observed.

©2008 Optical Society of America

## References and links

## 1. Introduction

Over the past decade there has been a renewed interest in the spectroscopic properties of the thulium ($Tm^{3+}$) ion, particularly in application areas such as optical communications, high power lasers, medicine and sensing. Of particular interest to this work is the application of thulium for optical amplification in the telecommunication S-band from 1460-1530 nm. Thulium doped fiber amplifiers (TDFAs) are amongst the leading candidates to bring the same effective means of optical amplification to the S-band as the erbium doped fiber amplifier (EDFA) has for the C-band and L-band.

To date, the only efficient TDFAs have been demonstrated in low phonon energy host materials (such as fluoride glasses) which are incompatible with existing silica based telecommunication infrastructure. For this technology to be a viable solution for S-band amplification efficient, robust and compatible amplifiers are required. TDFAs in a silica based host material are the preferred solution; however the spectroscopic properties of thulium result in silica being a poor host material for optical amplification. Various pumping techniques have been explored in the thulium system to improve the efficiency of the S-band transition [1-3], the most popular being the up-conversion pumping technique in which 1040-1060 nm pump photons are absorbed by the ground state and then the first excited state ($^3F_4$) to result in an excited ion in the upper amplifying $^3H_4$ manifold (see Fig 1.). Unfortunately, the energy level structure of the thulium supports a third absorption step from the upper amplifying manifold to a higher lying manifold ($^1G_4$) resulting in a quenching of the excited state population from the amplifying manifold [4], hence, research into other possible up-conversion pumping techniques is now under way. An alternate technique which is attracting considerable interest is co-doping thulium ions with other sensitising rare earth elements which can act to absorb and transfer energy. Ytterbium ($Yb^{3+}$) is seen as a promising candidate due to its significant absorption cross section and favorable energy level structure.

Studies of $Tm^{3+}/Yb^{3+}$ co-doped systems date back to the 1960s and 70s where Hewes et al. [5] and Ostermayer et al. [6] carried out extensive studies on the up-conversion characteristics of $Tm^{3+}/Yb^{3+}$ in $YF_3$ crystals. Researchers have since studied the properties of $Tm^{3+}/Yb^{3+}$ co-doped systems in a range of host materials [7-11], however limited work has been carried out on such systems in silica based glasses. Hanna et al. [12] reported on the up-conversion properties of $Tm^{3+}/Yb^{3+}$ co-doped silica glass under excitation at 1060 and 800-860 nm and found that the poor up-conversion efficiencies of the system were a result of the short excited state lifetimes of $Tm^{3+}$. However, recent work has demonstrated a 3-fold increase in the fluorescence lifetimes of the $^3H_4$ manifold through the incorporation of large amounts of $Al_2O_3$ into the silica glass network [13]. This, coupled with the development of high-powered semi-conductor laser sources around 980 nm, which enables the peak absorption of $Yb^{3+}$ to be optically pumped efficiently, suggests that an improvement in the up-conversion efficiencies in silica-based materials may now be realised.

In this work the up-conversion properties of $Tm^{3+}/Yb^{3+}$ co-doped alumino-silicate fibers under excitation at 980 nm are presented. The population mechanisms for the up-conversion processes are established and studied at three different $Tm^{3+}/Yb^{3+}$ concentration ratios. The lifetimes of the excited states of $Tm^{3+}$ and $Yb^{3+}$ are also reported and studied as a function of $Yb_2O_3$ concentration. By studying the population dynamics of the $Tm^{3+}/Yb^{3+}$ co-doped system in alumino-silicate glass, meaningful conclusions as to the system's potential to produce optical amplification in the S-band can be drawn. These conclusions also have important implications for $Tm^{3+}$ doped high power fiber lasers operating near 2 µm as the up-conversion mechanisms involved have been shown to reduce the performance of these devices [14-18].

## 2. Experimental details

Three $Tm^{3+}/Yb^{3+}$ co-doped alumino-silicate fibers were fabricated for this study using the MCVD and solution doping techniques. The core concentrations of the fibers are listed in Table 1 along with the $Tm^{3+}/Yb^{3+}$ concentration ratios.

Table 1: Core dopants of the $Tm^{3+}/Yb^{3+}$ co-doped silica fibers.

| Sample  | $Tm_2O_3$ (ppm) | $Yb_2O_3$ (ppm) | Ratio ($Tm^{3+}/Yb^{3+}$) | $Al_2O_3$ (mol %) |
|---------|-----------------|-----------------|---------------------------|-------------------|
| TmYb-1  | 100             | 430             | 1:4                       | 3.8               |
| TmYb-2  | 100             | 940             | 1:9                       | 3.8               |
| TmYb-3  | 100             | 1400            | 1:14                      | 3.2               |

The $Tm_2O_3$ and $Yb_2O_3$ concentrations given in Table 1 were estimated from the absorption peaks at 786 and 920 nm, respectively. The refractive index profiles of the fiber samples were measured, using a York S14 index profiler and used to obtain the $Al_2O_3$ concentrations since it has been shown that $Al_2O_3$ increases the refractive index of silica by $2.3 \times 10^{-3}$ per mol %; it was assumed that the concentration of $Tm^{3+}$ and $Yb^{3+}$ did not contribute significantly to the index difference. $Al_2O_3$ was used to modify the core region of the fiber, as it has been established that it has a lengthening effect on the fluorescence lifetime of the excited state manifolds. The $Al_2O_3$ also helps to distribute the rare earth ions homogeneously throughout the glass host. High $Tm_2O_3$ concentrations were avoided in this study as the cross relaxation processes between $Tm^{3+}$ ions may mask the energy transfer properties from the energy exchange between $Yb^{3+}$ and $Tm^{3+}$ ions. No other standard modifiers of silica, such as germanium, phosphorus or fluorine were used in the fabrication process.

The fluorescence intensity and lifetime measurements presented in the following sections were conducted by collecting the fluorescence immediately after the splice to the excitation source with a 0.5 NA aspheric lens transverse to the doped fibre. Sample lengths were kept to 50 mm to minimise the effects of amplified spontaneous emission and re-absorption. Table 2 summarises the experimental configuration for each energy manifold. Note: for the fluorescence lifetime measurements the pump laser decay time was around 50 ns.

Table 2: Experimental configuration for the measurement of fluorescence from the excited manifolds of $Tm^{3+}$ and $Yb^{3+}$.

| Energy manifold | Filter | Detector | Detector response time (µs) |
|---|---|---|---|
| $Tm^{3+}$ ($^3F_4$) | 1500 nm long pass | InGaAs (FGA20) | 13 |
| $Tm^{3+}$ ($^3H_4$) | 810 ± 10 nm | Photo-multiplier tube (R930) | 0.6 |
| $Tm^{3+}$ ($^1G_4$) | 475 ± 10 nm | Photo-multiplier tube (R930) | 0.6 |
| $Yb^{3+}$ ($^2F_{5/2}$) | 1030 ± 10 nm | InGaAs (FGA10) | 13 |

## 3 Fluorescence Intensity Measurements

The approach taken in this investigation was to use a double energy transfer process between $Yb^{3+}$ and $Tm^{3+}$ ions to further enhance the quantum efficiency of the S-band transition in $Tm^{3+}$. The proposed double energy transfer process has the added advantage of populating the upper amplifying $^3H_4$ manifold of $Tm^{3+}$ whilst depopulating the lower amplifying $^3F_4$ manifold, as shown in Fig. 1.

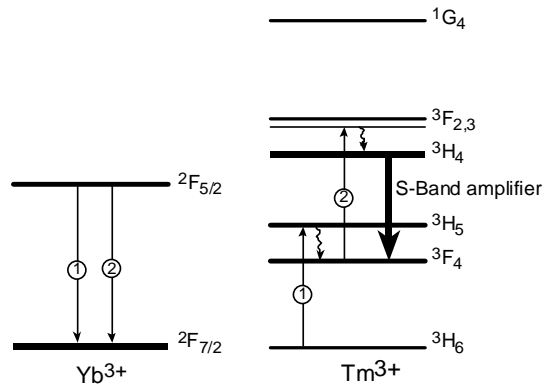

Fig. 1. Double energy transfer mechanism between $Tm^{3+}$ and $Yb^{3+}$ ions, under 980 nm excitation.

The double energy transfer mechanism involves the energy transfer of an excited ion in the $^2F_{5/2}$ manifold of $Yb^{3+}$ with a nearby ground state $Tm^{3+}$ ion, which excites the ground state $Tm^{3+}$ ion to the $^3H_5$ manifold. Due to the close proximity of the $^3F_4$ manifold, multi-phonon decay quickly relaxes any population in the $^3H_5$ manifold to the relatively long lived $^3F_4$ manifold. A second energy transfer from another excited $Yb^{3+}$ ion can then populate the $^3F_2$ and $^3F_3$ manifolds of $Tm^{3+}$. Again, multi-phonon decay quickly relaxes any population in the $^3F_2$ and $^3F_3$ manifolds to the $^3H_4$ manifold. The non-resonant nature of each energy transfer step necessitates the assistance of phonons. The energy mismatch for each up-conversion step is given below for $Tm^{3+}/Yb^{3+}$ in $YF_3$ [6]:

$Yb^{3+}$ ($^2F_{5/2}$) + $Tm^{3+}$ ($^3H_6$) → $Yb^{3+}$ ($^2F_{7/2}$) + $Tm^{3+}$ ($^3H_5$) $\Delta E \sim 1650$ cm$^{-1}$

$Yb^{3+}$ ($^2F_{5/2}$) + $Tm^{3+}$ ($^3F_4$) → $Yb^{3+}$ ($^2F_{7/2}$) + $Tm^{3+}$ ($^3F_{2,3}$) $\Delta E \sim 1000$ cm$^{-1}$

From the absorption spectra of the fiber samples used in this investigation these mismatches are estimated to be $1124 \pm 4$ and $822 \pm 33$ cm$^{-1}$, respectively. The reduction in the energy mismatches in silica glass may be attributed to the energy level broadening caused by the amorphous nature of the glass. The positive energy mismatch associated with these processes requires the emission of phonons to conserve energy.

### 3.1 Up-conversion pumping at 980 nm

When excited optically at 980 nm the fibers were found to emit blue luminescence that was clearly visible with the naked eye. The counter-propagating up-conversion luminescence spectrum for each fiber sample, between the wavelength range of 450 and 900 nm is shown in Fig. 2. It should be noted that the side fluorescence up-conversion spectrum was too weak to detect.

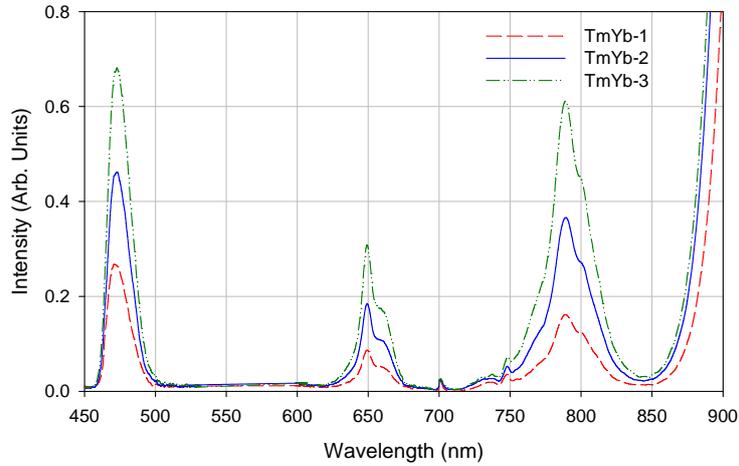

Fig. 2. Counter-propagating luminescence spectra of the $Tm^{3+}/Yb^{3+}$ co-doped fiber samples, under 980 nm excitation. Note: sample lengths were all kept to 20 cm and the incident pump power in each case was 128 mW.

The up-conversion luminescence spectra shows three distinct fluorescence bands centred around 475, 650 and 780 nm, which are attributed to the transitions from the $^1G_4 \rightarrow {}^3H_6$, $^1G_4 \rightarrow {}^3F_4$, and ($^1G_4 \rightarrow {}^3H_5$ & $^3H_4 \rightarrow {}^3H_6$) manifolds, respectively. Unfortunately, low fluorescence intensity levels prevented the counter propagating spectrum from the $^3H_4 \rightarrow {}^3F_4$ and $^3F_4 \rightarrow {}^3H_6$ transitions from being obtained. The three visible luminescence bands were found to increase with increasing $Yb^{3+}$ concentration and since these bands are not observed in $Tm^{3+}$ doped silica fibers under 980 nm excitation [12] it can be concluded that energy transfer is occurring between $Yb^{3+}$ and $Tm^{3+}$ ions.

To understand the origin of the luminescent bands the power dependencies of the up-conversion luminescence bands were studied as a function of the $Yb^{3+}$ excited state population for a range of incident pump powers. The number of pump photons, $n$, required to produce a up-converted photon in the $Tm^{3+}/Yb^{3+}$ co-doped system can be determined easily, as the up-conversion intensity is proportional to the power, $n$, on the density of excited atoms in the $^2F_{5/2}$ manifold of $Yb^{3+}$. Since the fluorescence intensity at 1060 nm is directly proportional to the density of excited atoms in the $^2F_{5/2}$ manifold of $Yb^{3+}$, the number of pump photons required for a particular up-conversion process is readily obtained from the slope of the up-conversion intensity versus the fluorescence intensity at 1060 nm.

### 3.1.1 $^3F_4$ manifold of $Tm^{3+}$

To establish the first energy transfer step, the fluorescence intensity at 1800 nm from the $^3F_4$ manifold of $Tm^{3+}$ was studied as a function of the fluorescence intensity at 1060 nm from $Yb^{3+}$. Figure 3 shows the log/log plot of the 1800 nm luminescence versus the 1060 nm luminescence for the three fibers.

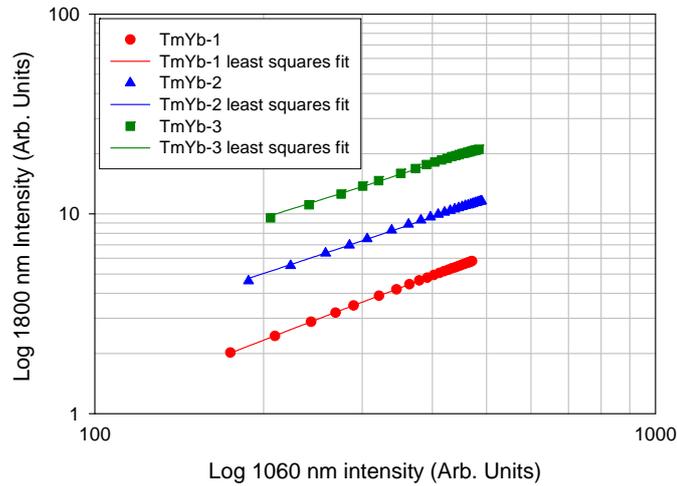

Fig. 3. Log/log plot of the 1800 nm luminescence from $Tm^{3+}$ vs. the 1060 nm luminescence from $Yb^{3+}$, for incident pump powers ranging from 3 - 108 mW. The 1800 nm measured data have been offset to aid comparison. Note: the errors associated with these measurements are not shown as they are smaller than the size of the individual data points in the plot.

The equation used to fit the measured data in Fig. 3 was obtained by solving the rate equations describing the excited state populations of the $^2F_{5/2}$ and $^3F_4$ manifolds. These rate equations are given below, along with the energy manifold labeling which is shown in Fig. 4, where $N_{Yi}$ represents the population of the $i^{th}$ $Yb^{3+}$ manifold and $N_{Tj}$ represents the population of the $j^{th}$ $Tm^{3+}$ manifold. It should be noted that the populations of the $^3H_5$ and $^3F_{2,3}$ manifolds of $Tm^{3+}$ have been ignored in this analysis due to the close spacing to the next lowest energy manifold; any population in these manifolds will decay non-radiatively to the longer lived $^3F_4$ and $^3H_4$ manifolds, respectively. $\tau_{Y1}$ represents the fluorescence lifetime of the $^2F_{5/2}$ manifold of $Yb^{3+}$, whilst $\tau_{Tj}$ represents the fluorescence lifetime of the $j^{th}$ excited manifold of $Tm^{3+}$. $W_i$ represents the energy transfer co-efficient describing the interaction between $Yb^{3+}$ and $Tm^{3+}$ ions for steps $i = 1$ to 3 and $\sigma_{Ti,j}$ represents the excited state absorption cross sections for transitions from the $i^{th}$ to $j^{th}$ manifold in $Tm^{3+}$. Finally, $\sigma_{Y01}$ is the absorption cross section of the ground to excited state transition of $Yb^{3+}$ when excited at 980 nm by pumping intensity $I$.

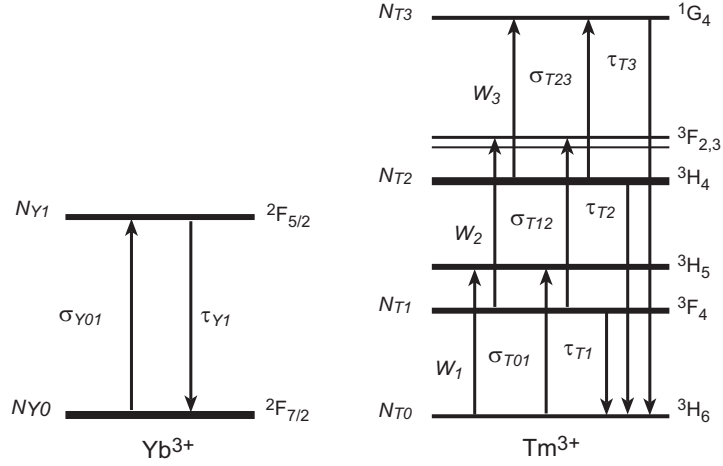

Fig. 4. Energy manifold labeling for the rate equation analysis of the $Tm^{3+}/Yb^{3+}$ co-doped system.

The rate equations describing the populations of the $^2F_{5/2}$ manifold of $Yb^{3+}$ and the $^3F_4$ manifold of $Tm^{3+}$ can be written as:

$$\frac{dN_{Y1}}{dt} = I\sigma_{Y01}N_{Y0} - \frac{N_{Y1}}{\tau_{Y1}}, \qquad (1)$$

and

$$\frac{dN_{T1}}{dt} = W_1 N_{Y1} N_{T0} - \frac{N_{T1}}{\tau_{T1}} - W_2 N_{T1} N_{Y1}, \qquad (2)$$

with the additional condition that:

$$N_{Y0} + N_{Y1} = c_Y, \qquad (3)$$

$$N_{T0} + N_{T1} + N_{T2} + N_{T3} = c_T, \qquad (4)$$

where $c_Y$ and $c_T$ represent the concentration of $Yb^{3+}$ and $Tm^{3+}$ ions, respectively.

It should be noted that Eq. (1) takes into account the assumption that the energy transfer up-conversion terms (i.e. $W_1 N_{Y1} N_{T0}$, $W_2 N_{Y1} N_{T1}$ and $W_3 N_{Y1} N_{T2}$) are significantly less than the spontaneous decay term $N_{Y1}/\tau_{Y1}$. This assumption is verified by fluorescence decay results from the $^2F_{5/2}$ manifold, reported in a later section. The second assumption made in this analysis is that, since only a small fraction of ground state $Tm^{3+}$ ions are excited by the up-conversion mechanisms, $N_{T0} \approx c_T$. The validity of this assumption is discussed further in the text. Equation (1) and (2) can be solved in the steady state, i.e. when $dN_{Y1}/dt$ and $dN_{T1}/dt = 0$, to obtain an expression for $N_{T1}$ as a function of $N_{Y1}$:

$$N_{T1} = \frac{W_1 N_{Y1} c_T}{\tau_{T1}^{-1} + W_2 N_{Y1}}. \qquad (5)$$

The solution shows that in the limit that $W_2 N_{Y1} \ll \tau_{T1}^{-1}$, a linear relationship exists between the population of the $^3F_4$ and $^2F_{5/2}$ manifolds. To test the validity of this solution, the measured data were fit to a linear expression in the form of $y = Ax + B$, where $A$ and $B$ were the fitting parameters. The excellent agreement, as shown in Fig. 3, between the fit and measured data over the entire pump power range for all fiber samples verifies that under 980 nm excitation, the $^3F_4$ manifold of $Tm^{3+}$, is populated by the energy transfer process:

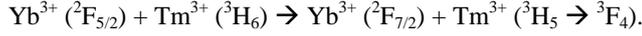

$Yb^{3+}$ ($^2F_{5/2}$) + $Tm^{3+}$ ($^3H_6$) → $Yb^{3+}$ ($^2F_{7/2}$) + $Tm^{3+}$ ($^3H_5$ → $^3F_4$).

It can also be concluded from the analysis that the energy transfer up-conversion rate, $W_2 N_{Yi}$, which depopulates the $^3F_4$ manifold is much less than the spontaneous decay, $\tau_{T1}^{-1}$.

### 3.1.2 *$^3H_4$ manifold of $Tm^{3+}$*

The second energy transfer step from the $^3F_4$ to the $^3F_{2,3}$ manifolds could not be determined from the luminescence at 810 nm in this sample set as, upon inspection of the $Tm^{3+}$ energy level diagram, see Fig. 1, luminescence from the $^1G_4$ → $^3H_5$ and $^3H_4$ → $^3H_6$ transitions both result in fluorescence bands around 810 nm. Since blue luminescence has been observed in these fiber samples it can be concluded that excited ions have been promoted to the $^1G_4$ manifold of $Tm^{3+}$. With this in mind, the 810 nm luminescence should show the properties associated with the populating mechanisms for both energy manifolds and hence an independent measurement of the $^3H_4$ manifold population cannot be obtained. This complication could be avoided by studying the luminescence properties of the 1480 nm transition; unfortunately the luminescence in this wavelength region was too weak to be detected, due to the low $Tm_2O_3$ concentrations and limited pump powers available. A short discussion of the 810 nm luminescence is given following the exploration of the $^1G_4$ manifold. This order has been chosen to aid understanding.

### 3.1.3 *$^1G_4$ manifold of $Tm^{3+}$*

The populating mechanism responsible for the blue luminescence from the $^1G_4$ manifold can be established by studying its dependence on the 1060 nm luminescence from $Yb^{3+}$. The fluorescence at 650 nm from the $^1G_4$ → $^3F_4$ transition could also be used to study the mechanism populating the $^1G_4$ manifold. However the branching ratio in silica glass for the 475 nm luminescence is 0.51 compared to 0.069 for the 650 nm luminescence [19], therefore the luminescence intensity of the 475 nm luminescence is at least 7 times stronger. Figure 5 shows the log/log plot of the 475 nm luminescence as a function of the $Yb^{3+}$ luminescence at 1060 nm for a range of input pump powers.

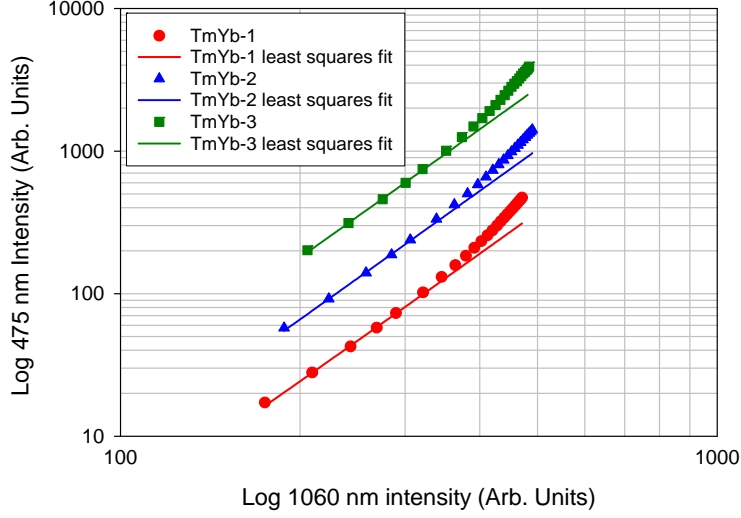

Fig. 5. Log/log plot of the 475 nm luminescence from $Tm^{3+}$ vs. the 1060 nm luminescence from $Yb^{3+}$, for incident pump powers ranging from 3 - 108 mW. The 475 nm measured data have been offset to aid comparison. Note: the errors associated with these measurements are not shown as they are smaller than the size of the individual data points presented in the plot. The fits to the data are discussed in the text.

The equation used to fit the measured data in Fig. 5 was determined from the solution to the rate equation describing the population of the $^1G_4$ manifold. Since the populating mechanism of the $^3H_4$ manifold could not be determined from the 810 nm luminescence data, it was assumed that a second energy transfer up-conversion process, involving the energy exchange between an excited $Yb^{3+}$ ion and an excited $Tm^{3+}$ ion in the $^3F_4$ manifold, populates the $^3H_4$ manifold under 980 nm excitation, as shown in Fig. 1. Based on this assumption the rate equation describing the population of the $^3H_4$ manifold and $^1G_4$ manifold of $Tm^{3+}$ can be written as:

$$\frac{dN_{T2}}{dt} = W_2 N_{Y1} N_{T1} - \frac{N_{T2}}{\tau_{T2}} - W_3 N_{T2} N_{Y1}, \quad (6)$$

and

$$\frac{dN_{T3}}{dt} = W_3 N_{Y1} N_{T2} - \frac{N_{T3}}{\tau_{T3}}. \quad (7)$$

Equation (7) assumes that the $^1G_4$ manifold is populated by a third energy transfer up-conversion process involving an excited $Yb^{3+}$ ion and an excited $Tm^{3+}$ ion in the $^3H_4$ manifold. This assumption was made on the basis of the significant body of work which attributes blue luminescence to the successive 3 step energy transfer up-conversion process [7, 9, 20-22].

By solving the rate equations for the respective manifolds in the steady state, expressions for the population of the $^3H_4$ and $^1G_4$ manifolds can be obtained as a function of $Yb^{3+}$ population. These solutions are:

$$N_{T2} = \frac{W_1 W_2 N_{Y1}^2 c_T \tau_{T1}}{\tau_{T2}^{-1} + W_3 N_{Y1}}, \quad (8)$$

and

$$N_{T3} = \frac{W_1 W_2 W_3 N_{Y1}^3 c_T \tau_{T1} \tau_{T3}}{\tau_{T2}^{-1} + W_3 N_{Y1}}. \quad (9)$$

Both expressions contain a saturation term, which plays a role only when the energy transfer up-conversion rate is comparable to the spontaneous decay rate from the $^3H_4$ manifold. It has been established that the energy transfer rate of the second energy transfer up-conversion (ETU) step ($W_2 N_{Y1}$) is much less than the spontaneous decay of the $^3F_4$ manifold; since the decay rate of the $^3H_4$ manifold is an order of magnitude greater than the $^3F_4$ manifold, it is suggested that the ETU rate of the third step from $^3H_4 \rightarrow {}^1G_4$ ($W_3 N_{Y1}$) is much less than the spontaneous rate from the $^3H_4$ manifold ($\tau_{T2}^{-1}$). Therefore the steady state solutions for the $^3H_4$ and $^1G_4$ manifolds can simplify to:

$$N_{T2} = W_1 W_2 N_{Y1}^2 c_T \tau_{T1} \tau_{T2}, \qquad (10)$$

$$N_{T3} = W_1 W_2 W_3 N_{Y1}^3 c_T \tau_{T1} \tau_{T2} \tau_{T3}. \qquad (11)$$

This results in the population of the $^3H_4$ manifold being dependent on the square of the $^2F_{5/2}$ population whilst, the $^1G_4$ manifold population is dependent on the cube of the $^2F_{5/2}$ population. Equation (11) was used to fit the measured data shown in Fig. 5. The fit describes the data accurately at low pump powers, but fails to describe the data at higher pump powers, for all three fiber samples. Figure 5 shows that the 475 nm luminescence continues to grow after the 1060 nm fluorescence begins to saturate. This behaviour cannot be described solely by the successive three step ETU process. The ETU process is inherently linked to the population of the excited states; hence as the $^2F_{5/2}$ manifold begins to saturate, so too will the other successive excited states of $Tm^{3+}$. The fact that the 475 nm luminescence continues to grow indicates that another populating mechanism is occurring within the co-doped system which is not simply dependent on the excited state populations, but also the incident pump power. The only energy transfer process which fulfils this criterion is excited state absorption (ESA), as it involves the energy exchange of a pump or fluorescent photon with an ion in an excited state. Since ESA requires only one accepting ion, the process is concentration independent and scales with incident pump or fluorescence power, which is an important difference when compared to ETU. Although ESA of $Tm^{3+}$ ions has been frequently reported in the literature [23-25], this has not been the case for $Tm^{3+}/Yb^{3+}$ co-doped systems. For ESA to occur in the $Tm^{3+}/Yb^{3+}$ co-doped system, excited $Tm^{3+}$ ions are required to absorb incident pump photons at 980 nm and/or fluorescing photons at 1060 nm. Of the possible ESA transitions which can occur in $Tm^{3+}$, only two are possible in the $Tm^{3+}/Yb^{3+}$ co-doped system under 980 nm excitation, namely the $^3F_4 \rightarrow {}^3F_{2,3}$ and $^3H_4 \rightarrow {}^1G_4$ transitions. The ESA cross sections for these two transitions have been calculated [26] and are shown in Fig. 6, with the $^3F_4 \rightarrow {}^3F_{2,3}$ transition exhibiting stronger absorption strengths compared to the $^3H_4 \rightarrow {}^1G_4$ transition. However, of most interest, is the location of these absorption peaks in regard to the energy available from the incident pump and fluorescing photons. Included in Fig. 6 is the emission spectrum of $Yb^{3+}$ under 980 nm excitation; the position of this fluorescence band indicates which ESA transition of $Tm^{3+}$ has the greatest spectral overlap with the energy available from the pump and fluorescing photons.

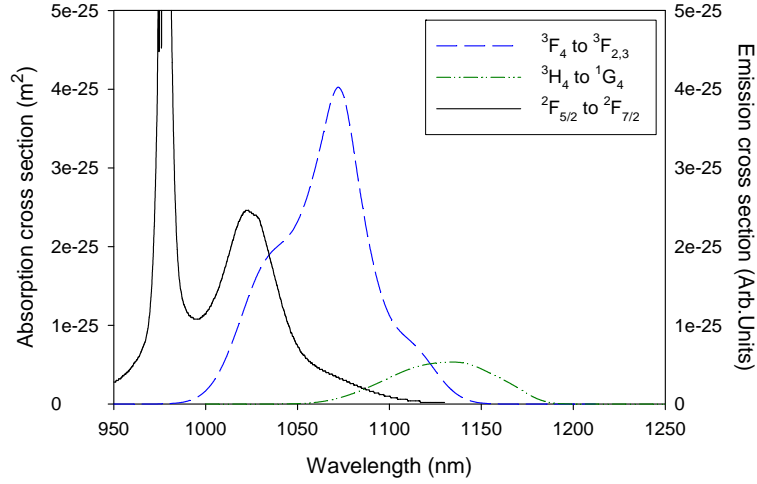

Fig. 6. Spectral overlap of the Yb$^{3+}$ fluorescence from the $^2F_{5/2} \rightarrow {}^2F_{7/2}$ transition with the calculated ESA transitions of Tm$^{3+}$ [26]. Note: the Yb$^{3+}$ fluorescence spectrum of the TmYb-1 sample was measured under 980 nm excitation.

The $^3F_4 \rightarrow {}^3F_{2,3}$ transition is found to have the greatest overlap with the Yb$^{3+}$ fluorescence but, more importantly, it is found to have a larger absorption cross section at the pump wavelength, 980 nm. The absorption cross section at the pump wavelength is the most critical parameter in this case as the number of incident pump photons is many orders of magnitude greater than the number of fluorescent photons. The ESA cross section of the $^3F_4 \rightarrow {}^3F_{2,3}$ transition at 980 nm in silica glass is estimated to be $5.2 \times 10^{-28}$ m$^2$, compared with $5.4 \times 10^{-36}$ m$^2$ for the $^3H_4 \rightarrow {}^1G_4$ transition. This comparison suggests that the $^3F_4 \rightarrow {}^3F_{2,3}$ transition is the most favourable ESA transition in the co-doped system under 980 nm excitation.

If the $^3F_4 \rightarrow {}^3F_{2,3}$ ESA transition is now considered in the populating dynamics of the Tm$^{3+}$/Yb$^{3+}$ co-doped system, the rate equation describing the population of the $^3H_4$ manifold would be given by Eq. (12); bearing in mind that the previous analysis determined that the energy transfer rates $W_1 N_{Y1}$, $W_2 N_{Y1}$, and $W_3 N_{Y1}$ are much less than the spontaneous decay rates from the $^3F_4$ and $^3H_4$ manifolds, respectively.

$$\frac{dN_{T2}}{dt} = I\sigma_{T12}N_{T1} + W_2 N_{Y1} N_{T1} - \frac{N_{T2}}{\tau_{T2}}. \quad (12)$$

Since ESA from the $^3H_4 \rightarrow {}^1G_4$ manifold is extremely unlikely under 980 nm pumping, the rate equation describing the population of the $^1G_4$ manifold would remain the same as that stated in Eq. (7). It should be noted that the inclusion of the ($^3F_4 \rightarrow {}^3F_{2,3}$) ESA term in the analysis also has implications on the rate equation describing the population of the $^3F_4$ manifold. However the linear dependence of the 1800 nm luminescence on the 1060 nm luminescence indicates that the ESA term is much less than the spontaneous decay rate from the $^3F_4$ manifold. Eqs. (12) and (7) can be solved in the steady state to obtain an expression for $N_{T3}$ as a function of $N_{Y1}$, namely:

$$N_{T3} = W_1 W_2 W_3 N_{Y1}^3 c_T \tau_{T1} \tau_{T2} \tau_{T3} + \frac{W_1 W_2 c_T \sigma_{T12} \tau_{T1} \tau_{T2} \tau_{T3} N_{Y1}^3}{\tau_{Y1} \sigma_{Y01}(c_Y - N_{Y1})}. \quad (13)$$

Equation (13) is similar to the expression obtained for the successive three step ETU process except for the additional term which results from the inclusion of ESA. The measured 475 nm versus 1060 nm luminescence data were then fitted with the new expression in the form of $y = Ax^3 + Bx^3/(1 - Cx)$, where $A$, $B$ and $C$ were the fitting parameters. The resulting fits are shown in Fig. 7 along with the $R^2$ value which provides an indication of the quality of the fit. The $A$, $B$ and $C$ fitting parameters are listed in Table 3 for each fiber sample along with their uncertainty.

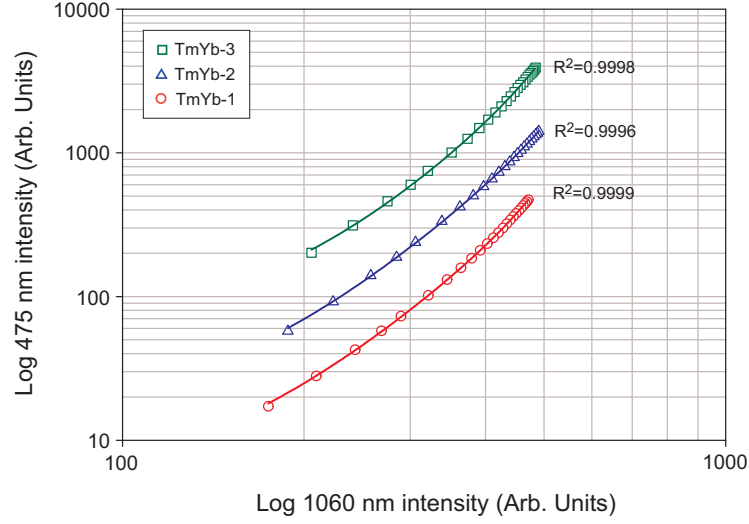

Fig. 7. Log/log plot of the 475 nm luminescence from $Tm^{3+}$ vs. the 1060 nm luminescence from $Yb^{3+}$, for incident pump powers ranging from 3 - 108 mW. The data were fitted with Eq. (13).

Table 3. Parameters obtained by fitting the steady state rate equation model to the 475 vs. 1060 nm luminescence data.

| Sample | Fitting parameters | | |
|---|---|---|---|
| | $A$ | $B$ | $C$ |
| TmYb-1 | $(1.8 \pm 0.1) \times 10^{-7}$ | $(5.6 \pm 0.5) \times 10^{-8}$ | $(1.67 \pm 0.03) \times 10^{-5}$ |
| TmYb-2 | $(1.6 \pm 0.1) \times 10^{-7}$ | $(6.0 \pm 0.7) \times 10^{-8}$ | $(1.52 \pm 0.03) \times 10^{-5}$ |
| TmYb-3 | $(1.5 \pm 0.2) \times 10^{-7}$ | $(6.1 \pm 0.9) \times 10^{-8}$ | $(1.61 \pm 0.04) \times 10^{-5}$ |

Although the physical parameters associated with the $A$, $B$ and $C$ terms cannot be obtained from the fitting parameters due to the coupling of several unknown parameters and proportionality constants, the uncertainty in the fitting parameters validates the model and the quality of the fit for all fiber samples. This establishes for the first time the $^3F_4 \rightarrow {^3F_{2,3}}$ ESA process as a populating mechanism in $Tm^{3+}/Yb^{3+}$ co-doped silica glasses under 980 nm excitation. It can therefore be concluded that the populating mechanisms involved in promoting $Tm^{3+}$ ions to the $^1G_4$ manifold in $Tm^{3+}/Yb^{3+}$ co-doped silica glass under 980 nm excitation are:

Step 1 - $Yb^{3+} ({^2F_{5/2}}) + Tm^{3+} ({^3H_6}) \rightarrow Yb^{3+} ({^2F_{7/2}}) + Tm^{3+} ({^3H_5} \rightarrow {^3F_4})$
Step 2 - $Yb^{3+} ({^2F_{5/2}}) + Tm^{3+} ({^3F_4}) \rightarrow Yb^{3+} ({^2F_{7/2}}) + Tm^{3+} ({^3F_{2,3}} \rightarrow {^3H_4})$ &
(980 nm photons) + $Tm^{3+} ({^3F_4}) \rightarrow Tm^{3+} ({^3F_{2,3}} \rightarrow {^3H_4})$
Step 3 - $Yb^{3+} ({^2F_{5/2}}) + Tm^{3+} ({^3H_4}) \rightarrow Yb^{3+} ({^2F_{7/2}}) + Tm^{3+} ({^1G_4})$

Although the quantum efficiency of the S-band transition cannot be quantified in this sample set, the spectroscopic study of the system has established two energy transfer processes that act to populate the upper amplifying $^3H_4$ manifold whilst depopulating the lower amplifying $^3F_4$ manifold, under 980 nm excitation.

*3.1.4 810 nm luminescence*

As discussed previously, the up-conversion luminescence at 810 nm is attributed to the presence of two overlapping luminescence bands, one from each of the $^1G_4$ and $^3H_4$ manifolds. In this case, the luminescence at 810 nm should exhibit the characteristics of both excited manifolds with the $^3H_4$ manifold dominating over the $^1G_4$ manifold at low excitation powers. Figure 8 shows the log/log plot of the 810 nm luminescence as a function of the $Yb^{3+}$ luminescence at 1060 nm over a range of input pump powers for each fiber sample.

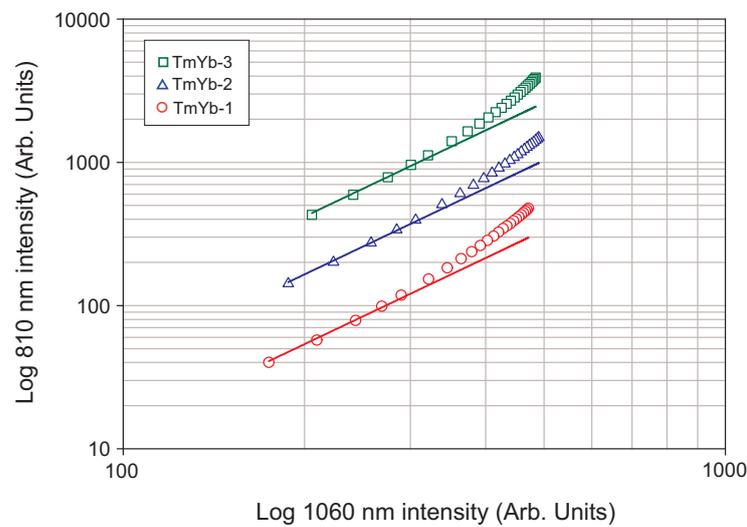

Fig. 8. Log/log plot of the 810 nm luminescence from $Tm^{3+}$ vs. the 1060 nm luminescence from $Yb^{3+}$, for incident pump powers ranging from 3 - 108 mW. The 810 nm measured data have been offset to aid comparison. The solid lines represent the fit of a standard quadratic equation to the measured data.

The non-linear nature of the log-log plot suggests the presence of more than one up-conversion process. At low pump powers the measured data exhibits a slope of 2, as seen in Fig. 8, whilst at high pump powers > 50 mW the slope exceeds 3. The non-linear behaviour is greatest in the TmYb-3 sample which has the highest $^1G_4$ manifold population. This provides further evidence of the two overlapping transitions from the $^1G_4$ and $^3H_4$ manifold and is consistent with the rate equation model proposed here. Unfortunately, the large number of unknown parameters prevents the rate equation model from being fitted to the measured data with any degree of certainty. A more accurate account of the population dynamics involved in this luminescence band can be obtained by studying the fluorescence decay characteristics after the pump excitation has been removed.

## 4 Fluorescence Lifetime Measurements

The final stage in the spectroscopic study of the $Tm^{3+}/Yb^{3+}$ co-doped system was to investigate the fluorescence lifetimes of the excited states of $Tm^{3+}$ and $Yb^{3+}$. The fluorescence lifetimes of the $^2F_{5/2}$, $^3F_4$, and $^3H_4$ manifolds were studied under direct excitation at the appropriate wavelength. In addition, the $^3F_4$, $^3H_4$ and $^1G_4$ manifolds of $Tm^{3+}$ were studied under in-direct pumping at 980 nm.

**4.1 Direct pumping**

*4.1.1 $^2F_{5/2}$ manifold of $Yb^{3+}$*

The decay characteristics of the $^2F_{5/2}$ manifold of $Yb^{3+}$ were studied in the $Tm^{3+}/Yb^{3+}$ -co-doped fibers by directly exciting $Yb^{3+}$ ions to the $^2F_{5/2}$ manifold at 980 nm. Figure 9 shows the normalised measured decay waveform from the $^2F_{5/2}$ manifold under 980 nm excitation. The fluorescence decay was measured using 50 ms pulses at a repetition rate of 10 Hz.

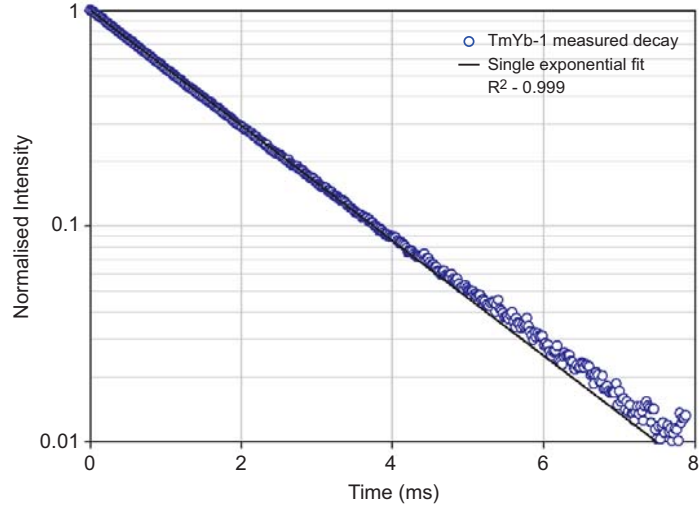

Fig. 9. Semi-log plot of the normalised 1060 ± 10 nm fluorescence decay waveform from the $^2F_{5/2}$ manifold of sample TmYb-1 under 980 nm pulsed excitation.

The solid line shown in Fig. 9 was obtained by fitting a single exponential function to the measured data. The single exponential fit is in excellent agreement with the measured waveform. The 1/e lifetimes obtained from the single exponential fit are listed in Table 4 for the three $Tm^{3+}/Yb^{3+}$ -co-doped fibers.

Table 4: Fluorescence lifetimes of the $^2F_{5/2}$ manifold under 980 nm excitation with an incident pump power of 9.3 mW.

| Sample | $^2F_{5/2}$ lifetime (µs) |
|--------|---------------------------|
| TmYb-1 | 826.0 ± 0.6 |
| TmYb-2 | 832.0 ± 0.6 |
| TmYb-3 | 823.0 ± 0.5 |

The measured lifetimes of the three alumino-silicate samples reported in Table 4 are consistent with those reported for $Yb^{3+}$-doped alumino-silicate glass [27], which provides strong evidence that the assumption used in the rate equation analysis that the energy transfer terms $W_1N_{Y1}N_{T0}$, $W_2N_{Y1}N_{T1}$, and $W_3N_{Y1}N_{T2}$ are much less than $N_{Y1}/\tau_{Y1}$ is valid. This is also supported by the lack of lifetime dependence on the $Yb_2O_3$ concentration.

*4.1.2 $^3F_4$ and $^3H_4$ manifolds of $Tm^{3+}$*

The decay characteristics of the $^3F_4$ and $^3H_4$ manifolds of $Tm^{3+}$ were studied by exciting the manifolds directly at 1586 and 780 nm, respectively. The luminescence decay from both excited manifolds was characterised by a single exponential function. A more rigorous treatment of the decay characteristics of these manifolds in $Tm^{3+}$ doped silica glass has been done in [28]. In that work the decay from the $^3H_4$ and $^3F_4$ manifolds under direct excitation was shown to exhibit a degree of non-exponentiality which was attributed to the distribution of possible multi-phonon decay rates in the glass matrix. The measured decay waveforms in this work were found to exhibit similar degrees of non exponentiality than those observed in $Tm^{3+}$ doped silica fibers. Hence, the non exponential nature of the decay can be attributed to the host matrix rather than possible energy transfer processes between $Yb^{3+}$ and $Tm^{3+}$ ions. The single exponential fits applied in this work allow comparisons to be made between each fiber sample and aid comparison between the measured lifetimes reported the literature. The 1/e lifetime of the $^3F_4$ and $^3H_4$ manifolds are listed in Table 5, respectively.

Table 5: Fluorescence lifetimes of the $^3F_4$ and $^3H_4$ manifolds under direct excitation at 1586 and 780 nm, respectively. Where $\tau_{1/e}$ represents the lifetime obtained from the single exponential fit. Note: 30 μs pulses at 10 Hz were used for direct excitation of the $^3F_4$ manifold, whilst 3 μs pulses at a repetition rate of 10 Hz were used to excite the $^3H_4$ manifold directly.

| Sample | $^3F_4$ lifetime $\tau_{1/e}$ (μs) | $^3H_4$ lifetime $\tau_{1/e}$ (μs) |
|---|---|---|
| TmYb-1 | 424 ± 2 | 20.0 ± 0.1 |
| TmYb-2 | 423 ± 2 | 20.1 ± 0.1 |
| TmYb-3 | 377 ± 2 | 18.7 ± 0.1 |

The fluorescence lifetimes of both excited manifolds are consistent with those reported in $Tm^{3+}$-doped alumino-silicate fibers [29]. Sample TmYb-3 did however exhibit a considerable reduction when compared to the other $Tm^{3+}/Yb^{3+}$ samples. This reduction is attributed to the glass host rather than potential energy back transfer processes and cross relaxation effects. Previous work on singly doped $Tm^{3+}$ alumino-silicate has shown that the lifetime of the $^3F_4$ and $^3H_4$ manifolds are dependent on the amount of $Al_2O_3$ present in the fiber core [28]. In that work the fluorescence lifetime of both excited manifolds were found to decrease with decreasing amounts of $Al_2O_3$. This is consistent with results presented in Table 5 as the TmYb-3 sample contains ~ 15% less $Al_2O_3$ in the core when compared to the other two co-doped samples.

**4.2 In-direct pumping**

The in-direct excitation of the $^3F_4$, $^3H_4$ and $^1G_4$ manifolds of $Tm^{3+}$ has been demonstrated in the previous section under continuous wave excitation at 980 nm. In this section, the fluorescence decay characteristics of the excited manifolds are studied under pulsed excitation. The time dependent rate equation model established in the previous section for continuous wave pumping can be carried over into the fluorescence decay analysis to describe the excited state population over time, allowing the validity of the model to be tested under two energy excitation regimes.

*4.2.1 $^3F_4$ manifold of $Tm^{3+}$*

The fluorescence decay of the $^3F_4$ manifold under in-direct excitation at 980 nm is shown in Fig. 10 for the TmYb-3 sample. The decay characteristics of the $^3F_4$ manifold showed considerable differences to those obtained under direct excitation.

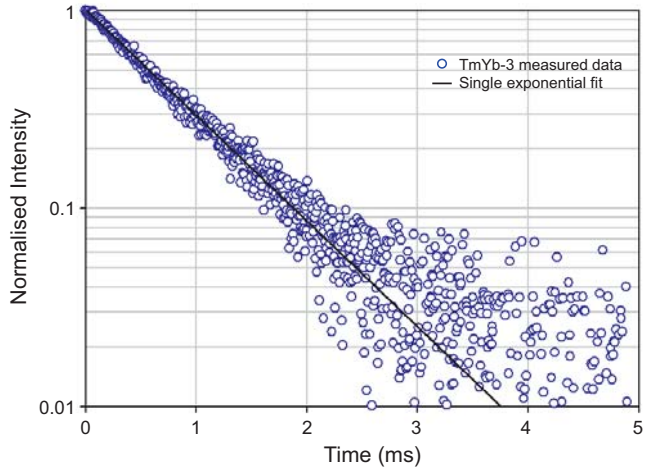

Fig. 10. Semi-log plot of the normalised fluorescence decay waveform from the $^3F_4$ manifold of sample TmYb-3 under 980 nm excitation.

The fluorescence decay was described well by a single exponential function, as seen in Fig. 10. The characteristic lifetime of the manifold was found to increase by a factor of two on average, when compared to the lifetimes obtained under direct pumping and were comparable to the fluorescence lifetime of the $^2F_{5/2}$ manifold of $Yb^{3+}$ (see Table 6).

Table 6: Fluorescence lifetime of the $^3F_4$ manifold under in-direct excitation at 980 nm. Note: 50 ms pulses at 10 Hz were used to excite the $^3F_4$ manifold in-directly. The fluorescence lifetimes of the $^3F_4$ and $^2F_{5/2}$ manifolds under direct excitation at 1586 and 980 nm, respectively, are shown for comparison.

| Sample | In-Direct pumping (980 nm) $\tau_{1/e}$ (μs) | Direct pumping (1586 nm) $\tau_{1/e}$ (μs) | $^2F_{5/2}$ manifold (980 nm) $\tau_{1/e}$ (μs) |
|---|---|---|---|
| TmYb-1 | 781 ± 10 | 424 ± 2 | 826.0 ± 0.6 |
| TmYb-2 | 845 ± 20 | 423 ± 2 | 832.0 ± 0.6 |
| TmYb-3 | 812 ± 10 | 377 ± 2 | 823.0 ± 0.5 |

The considerably large error associated with the single exponential fit was a result of the poor signal to noise ratio of the 1800 nm luminescence under in-direct pumping, rather than the inability of the fit to describe the waveform accurately. The effective doubling of the fluorescence lifetime of the $^3F_4$ manifold under in-direct pumping provides strong evidence that efficient energy transfer is occurring from the $^2F_{5/2}$ manifold of $Yb^{3+}$ to the $^3F_4$ manifold of $Tm^{3+}$, as the decay of the $^3F_4$ manifold of $Tm^{3+}$ is being dictated by the longer lived fluorescence lifetime of the $Yb^{3+}$ excited state. It also reaffirms the conclusions draw in Section 3.1.1 that the decay rates associate with ETU and ESA are much less than the spontaneous decay rate of the $^3F_4$ manifold.

### 4.2.2 $^1G_4$ manifold of $Tm^{3+}$

The in-direct excitation of the $^1G_4$ manifold of $Tm^{3+}$ can be achieved by the series of successive energy transfer processes as discussed in the previous section. The fluorescence decay from the $^1G_4$ manifold under in-direct pumping at 980 nm is shown in Fig. 11 for the TmYb-1 sample. The measured waveforms for all samples were described adequately by a single exponential fit. The characteristic lifetimes obtained from the single exponential fits are listed in Table 7 for the three co-doped samples.

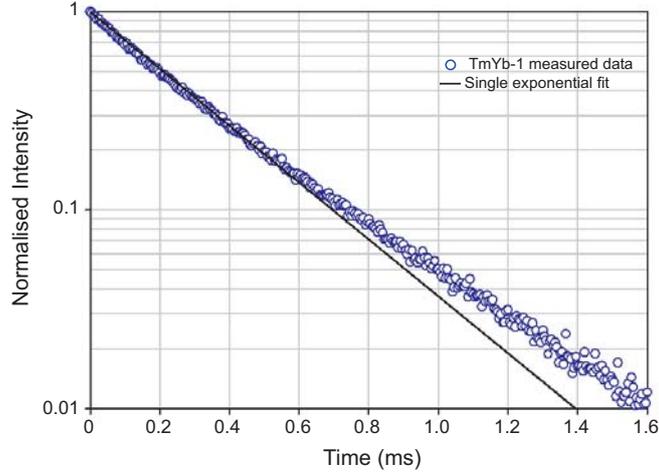

Fig. 11. Semi-log plot of the normalised fluorescence decay from the $^1G_4$ manifold under 980 nm excitation for sample TmYb-1. Note: decay waveforms were recorded using 50 ms pulses at a repetition rate of 10 Hz.

Table 7: Fluorescence lifetime of the $^1G_4$ manifold under in-direct excitation at 980 nm. Note: 50 ms pulses at 10 Hz were used to excite the $^3F_4$ manifold.

| Sample | $\tau_{1/e}$ (μs) |
|--------|-------------------|
| TmYb-1 | 302.4 ± 0.7 |
| TmYb-2 | 304.3 ± 0.4 |
| TmYb-3 | 298.2 ± 0.5 |

The single exponential nature of the $^1G_4$ decay provides information regarding the mechanisms dominating the $^1G_4$ manifold population after the pump excitation has been removed. The time dependent rate equation describing the population of the $^1G_4$ manifold is given by:

$$\frac{dN_{T3}(t)}{dt} = W_3 N_{Y1}(t) N_{T2}(t) - \frac{N_{T3}}{\tau_{T3}} \quad (14)$$

Although the solution to Eq. 14 cannot be obtained without the knowledge of $N_{T2}(t)$, it is clear that the solution will contain several exponential components each with their own amplitude and characteristic time constant, resulting in a non-exponential decay waveform. However, the experimentally observed fluorescence decay waveforms were sufficiently single exponential in nature with characteristic time constants consistent with the $^1G_4$ manifold lifetime reported in $Tm^{3+}$-doped silica fibers [30, 31]. It can therefore be concluded that the energy transfer rate into the $^1G_4$ manifold is much less than the spontaneous decay. Therefore, the $^1G_4$ manifold decay can be described by the single exponential function:

$$N_{T3}(t) = N_{T3}(0) \exp\left(-\frac{t}{\tau_{T3}}\right) \quad (15)$$

This is not a surprising result; the lifetime of the $^3H_4$ manifold is relatively short and hence the likelihood of an excited ($^3H_4$) $Tm^{3+}$ ion interacting with an excited ($^2F_{5/2}$) $Yb^{3+}$ ion after the pump excitation has been removed is extremely low. The other important point is that the lifetime of the $^1G_4$ manifold remains constant over the $Yb_2O_3$ concentration range studied here, providing further evidence that the energy transfer rate is negligible compared to the spontaneous decay. This also suggests that there is negligible energy back transfer from the $^1G_4$ manifold of $Tm^{3+}$ to the $^2F_{5/2}$ manifold of $Yb^{3+}$.

*4.2.3 810 nm luminescence*

The fluorescence decay properties of the 810 nm luminescence were studied in an effort to verify the existence of the two overlapping transitions from the $^3H_4$ and $^1G_4$ manifolds. The fluorescence decay of the $^3H_4$ manifold under in-direct pumping at 980 nm is shown in Fig. 12 for the TmYb-1 sample.

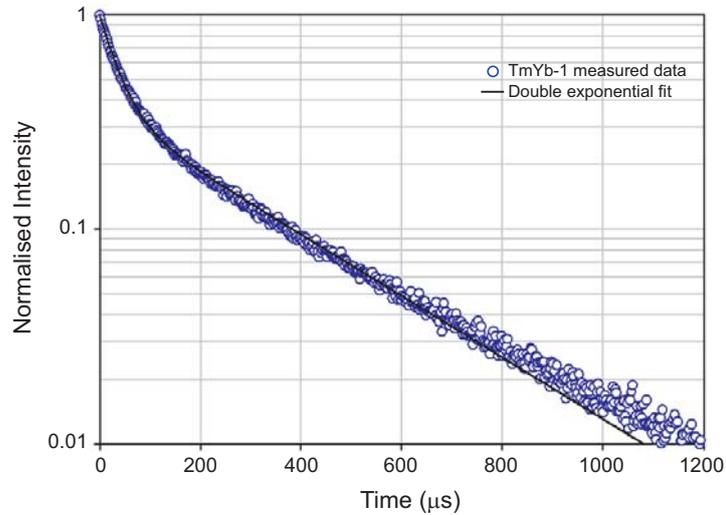

Fig. 12. Semi-log plot of the normalised fluorescence decay at 810 nm under 980 nm excitation for sample TmYb-1. Note: decay waveforms were recorded using 50 ms pulses at a repetition rate of 10 Hz. The double exponential fit is discussed in the text.

The decay waveforms at 810 nm were described accurately by a double exponential function in the form of $y = (1-A) \exp(-x/B) + A \exp(-x/C)$, where $A$ was the fitting parameter describing the amplitude of the second exponential and $B$ and $C$ were the fitting parameters used to obtain the two characteristic lifetimes. The two characteristic lifetimes obtained from the double exponential fit are listed in Table 8 for the three co-doped fibers.

Table 8: Amplitude and characteristic lifetimes obtained from the double exponential fit of the 810 nm fluorescence decay for all three samples under in-direct pumping at 980 nm.

| Sample | Amplitude '$A$' | Fast decay '$B$' (µs) | Slow decay '$C$' (µs) |
| --- | --- | --- | --- |
| TmYb-1 | $0.355 \pm 0.001$ | $37.1 \pm 0.2$ | $302 \pm 1$ |
| TmYb-2 | $0.406 \pm 0.002$ | $36.6 \pm 0.2$ | $307 \pm 1$ |
| TmYb-3 | $0.434 \pm 0.002$ | $30.9 \pm 0.2$ | $301 \pm 1$ |

The cause of the double exponential decay is attributed to the presence of two overlapping transitions in the $810 \pm 10$ nm window. It has been proposed throughout this paper that the $^1G_4 \rightarrow {}^3H_5$ transition spectrally overlaps the $^3H_4 \rightarrow {}^3H_6$ transition. The fluorescence decay in such a case would exhibit the characteristics from each energy manifold. It was established in the previous section, that the $^1G_4$ manifold decays in a single exponential form with a characteristic lifetime around ~ 300 µs. Assuming the $^3H_4$ manifold exhibits similar characteristics to the $^1G_4$ manifold, i.e. the energy transfer rates which populate and depopulate the manifold are much less than the spontaneous decay, the resultant decay would be a double exponential with two characteristic time constants equivalent to the lifetimes of the $^3H_4$ and $^1G_4$ manifolds. The slow component of the 810 nm luminescence decay is in excellent agreement with the measured $^1G_4$ manifold lifetime, whilst the fast component of the 810 nm decay is of the order of the $^3H_4$ manifold lifetime. The $^1G_4$ manifold contribution to the 810 nm luminescence may explain the increase in the amplitude of the slow component of the decay with increasing $Yb_2O_3$ concentration. The $^1G_4$ luminescence intensity was found to increase at a greater rate than the $^3H_4$ luminescence with increasing $Yb_2O_3$ concentration (refer to Fig. 2), and hence the contribution the $^1G_4$ manifold makes to the overall luminescence at 810 nm increases with increasing $Yb_2O_3$ concentration.

**Summary**

The spectroscopic study of the $Tm^{3+}/Yb^{3+}$-co-doped system in alumino-silicate glass identified up-conversion luminescence in the visible and near infra-red regions under 980 nm excitation. The steady state rate equation analysis established two energy transfer processes capable of depleting the $^3F_4$ manifold and populating the $^3H_4$ manifold. A double energy transfer up-conversion process and an excited state absorption process were identified as populating mechanisms in the co-doped system under 980 nm excitation. These two processes are the key to the $Tm^{3+}/Yb^{3+}$-co-doped system becoming an efficient S-band amplifying source. The other significant result to come from the analysis was that there was little evidence of energy back transfer from $Tm^{3+}$ to $Yb^{3+}$ ions under the direct pumping of the $^3F_4$ and $^3H_4$ manifolds.

A drawback of the $Tm^{3+}/Yb^{3+}$-co-doped system is the presence of a third energy transfer up-conversion process which transfers population from the upper amplifying $^3H_4$ manifold to the $^1G_4$ manifold. The rate of quenching of the $^3H_4$ manifold has not been identified in this analysis; however from the fluorescence lifetime results it is considered to be significantly less than the ~ 3257 $s^{-1}$ decay rate of the $^1G_4$ manifold.

**Acknowledgments**

This work was supported by the Australian Research Council, and Centre National de la Recherche Scientifique, in France.